%% file: paper.tex
\documentclass[acmscmall,authorversion=true, noacm=true]{acmart}
        
\usepackage{todonotes}
\usepackage{cleveref}
\usepackage{tikz}
\usetikzlibrary{shapes,arrows}
\usetikzlibrary{positioning}
\usetikzlibrary{fit}
\usetikzlibrary{decorations.pathreplacing}
\usetikzlibrary{tikzmark}
\usepackage{pgfplots}
\usepackage{pifont}
\usepackage{listings}
\usepackage{courier}
\newcommand{\cmark}{\ding{51}}%
\newcommand{\xmark}{\ding{55}}%
\usepackage[acronym]{glossaries}
\input{glossary}

\AtBeginDocument{%
  \providecommand\BibTeX{{%
    \normalfont B\kern-0.5em{\scshape i\kern-0.25em b}\kern-0.8em\TeX}}}

\setcopyright{none}
\copyrightyear{2023}

\begin{document}

\title{Lowering the Entry Bar to HPC-Scale Uncertainty Quantification}

\author{Linus Seelinger}
\orcid{0000-0001-8632-8493}
\affiliation{%
  \department{Institute for Applied Mathematics}
  \institution{Heidelberg University}
  \city{Heidelberg}
  \country{Germany}}
\email{mail@linusseelinger.de}

\author{Anne Reinarz}
\orcid{0000-0003-1787-7637}
\affiliation{%
  \department{Department of Computer Science}
  \institution{Durham University}
  \city{Durham}
  \country{United Kingdom}
}
\email{anne.k.reinarz@durham.ac.uk}

\author{Jean Bénézech}
\orcid{0000-0002-2989-3856s}
\affiliation{%
  \department{Department of Mechanical Engineering}
  \institution{University of Bath}
  \city{Bath}
  \country{United Kingdom}
}
\email{jb3285@bath.ac.uk}

\author{Mikkel Bue Lykkegaard}
\orcid{0000-0002-0932-9668}
\affiliation{%
  \institution{digiLab}
  \city{Exeter}
  \country{United Kingdom}
}
\email{mikkel@digilab.co.uk}

\author{Lorenzo Tamellini}
\orcid{0000-0002-8008-0359}
\affiliation{%
  \department{Institute of Applied Mathematics and Information Technologies}
  \institution{National Research Council}
  \city{Pavia}
  \country{Italy}
}
\email{tamellini@imati.cnr.it}

\author{Robert Scheichl}
\orcid{0000-0001-8493-4393}
\affiliation{%
  \department{Institute for Applied Mathematics}
  \institution{Heidelberg University}
  \city{Heidelberg}
  \country{Germany}
}
\email{r.scheichl@uni-heidelberg.de}

\renewcommand{\shortauthors}{Seelinger, et al.}

\begin{abstract}

Treating uncertainties in models is essential in many fields of science and engineering.
Uncertainty quantification (UQ) on complex and computationally costly numerical models necessitates a combination of efficient model solvers, advanced UQ methods and HPC-scale resources. The resulting technical complexities as well as lack of separation of concerns between UQ and model experts is holding back many interesting UQ applications.

The aim of this paper is to close the gap between advanced UQ methods and advanced models by removing the hurdle of complex software stack integration, which in turn will offer a straightforward way to scale even prototype-grade UQ applications to high-performance resources.

We achieve this goal by introducing a parallel software architecture based on UM-Bridge, a universal interface for linking UQ and models. We present three realistic applications from different areas of science and engineering, scaling from single machines to large clusters on the Google Cloud Platform.
\end{abstract}


\begin{teaserfigure}
\centering
  \includegraphics[width=0.8\textwidth]{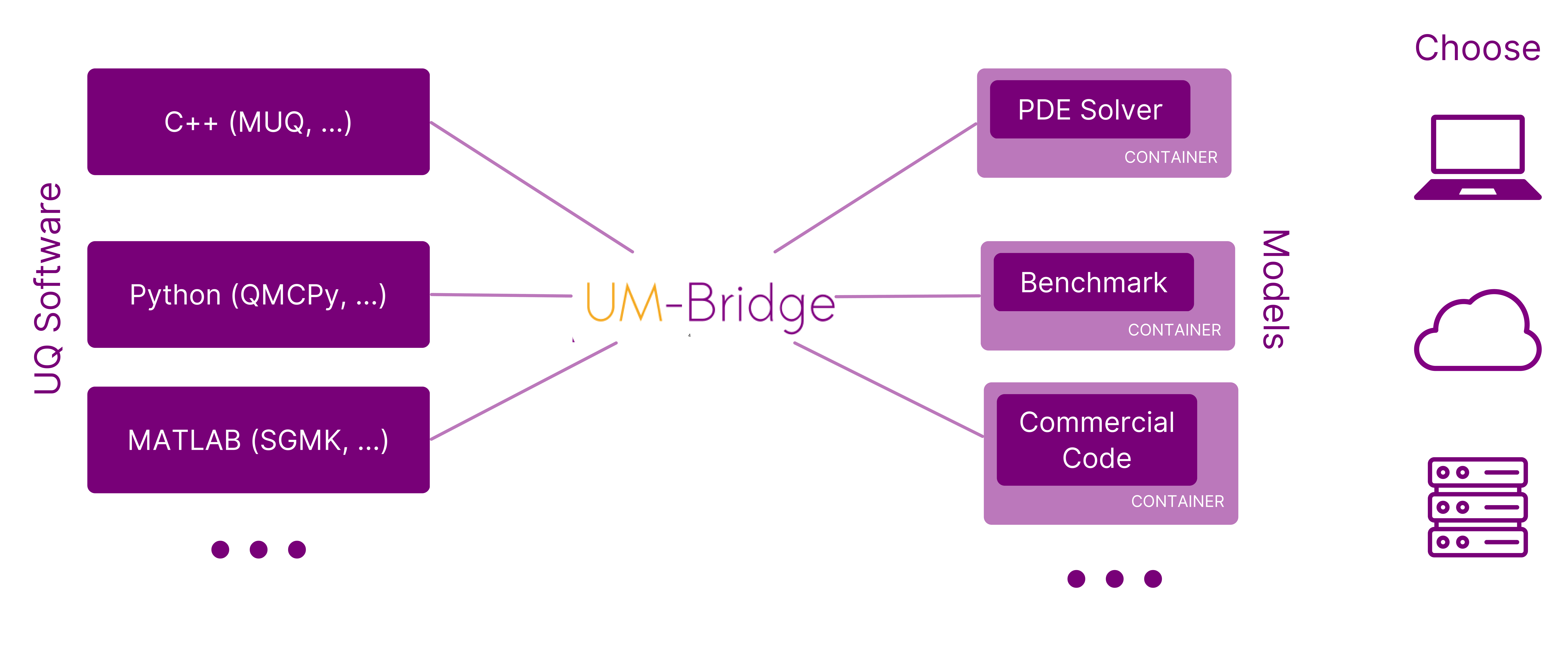}
  \caption{Coupling between uncertainty quantification and model codes using the UM-Bridge interface. UM-Bridge scales from small prototypes up to HPC-grade clusters or the cloud. }
  \label{fig:teaser}
\end{teaserfigure}

\maketitle

\input{introduction}

\input{umbridge}

\input{scalability}

\input{applications}

\input{conclusion}

\begin{acks}

We would like to acknowledge support by Cristian Mezzanotte and the Google Cloud Platform research credits program.
Lorenzo Tamellini has been supported by the PRIN 2017 project 201752HKH8 
``Numerical Analysis for Full and Reduced Order Methods for the efficient and accurate 
solution of complex systems governed by Partial Differential Equations (NA-FROM-PDEs)'' 
and by the  Research program CN00000013 ``National Centre for HPC, Big Data and Quantum 
Computing -- Spoke 6 - Multiscale Modelling \& Engineering Applications''.
The authors acknowledge support by the state of Baden-Württemberg through bwHPC
and the German Research Foundation (DFG) through grant INST 35/1597-1 FUGG. 

\end{acks}

\bibliographystyle{ACM-Reference-Format}
\bibliography{bibliography}

\end{document}

%% file: glossary.tex
\newacronym{UQ}{UQ}{Uncertainty Quantification}
\newacronym{MC}{MC}{Monte Carlo}
\newacronym{MCMC}{MCMC}{Markov chain Monte Carlo}
\newacronym{MLMCMC}{MLMCMC}{Multilevel Markov chain Monte Carlo}
\newacronym{MLMC}{MLMC}{Multilevel Monte Carlo}
\newacronym{HPC}{HPC}{High Performance Computing}
\newacronym{DUNE}{DUNE}{Distributed and Unified Numerics Environment}
\newacronym{FE}{FE}{finite element}
\newacronym{PDE}{PDE}{partial differential equation}
\newacronym{MUQ}{MUQ}{MIT UQ Library}
\newacronym{UMBridge}{UM-Bridge}{UQ and Modeling Bridge}
\newacronym{HTTP}{HTTP}{Hypertext Transfer Protocol}
\newacronym{MPI}{MPI}{Message Passing Interface}
\newacronym{MAP}{MAP}{maximum a posteriori probability}
\newacronym{pdf}{PDF}{probability density function}
\newacronym{sgmk}{SGMK}{Sparse Grids Matlab Kit}
\newacronym{GKE}{GKE}{Google Kubernetes Engine}
\newacronym{SMT}{SMT}{Simultaneous Multithreading}
\newacronym{QMC}{QMC}{Quasi-Monte Carlo}
\newacronym{GCP}{GCP}{Google Cloud Platform}
\newacronym{GP}{GP}{Gaussian Process}

%% file: introduction.tex
\section{Introduction}

The overarching goal of \gls{UQ} is to quantitatively assess the effect of uncertain parameters or data on a given system. For a given deterministic mathematical model, uncertainties in model parameters may be incorporated by treating the parameters as random variables of some distribution in the sense of probability theory. This covers both the aleatoric case, where model parameters are believed to be actually random in nature, and the epistemic case, where a specific real-world value is assumed to exist but is only known up to some degree of certainty.

A forward \gls{UQ} problem is then to determine the distribution of model predictions implied by the supposed distribution of model parameters. Inverse \gls{UQ} problems on the other hand start from real-world observations, often affected by uncertainties themselves, and infer or update the distribution of underlying model parameters that explain the observations.

There is a wide variety of methods for the numerical solution of both forward and inverse \gls{UQ} problems. Some examples for forward problems are \gls{MC} \cite{MC}, \gls{QMC} \cite{QMCReview}, \gls{MLMC} \cite{MLMC}, stochastic collocation \cite{SC_Survey}, and stochastic Galerkin \cite{GhanemSpanos_StochasticGalerkin,XiaKarniadakis_StochasticGalerkin}. Methods for inverse problems include the \gls{MCMC} family of methods \cite{MHMCMC,DILI,pCNOrig,pCNGeneralized,AMMCMC,HamiltonianMCMC,NUTS} and their multilevel siblings \cite{MLMCMCRevised,SchwabMLMCMC,MultilevelDILI,Lykkegaard_MLDA,lykkegaard_multilevel_2023}, importance sampling \cite{ImportanceSampling}, stochastic collocation for inverse problems \cite{Marzouk_StochasticCollocation}, and optimization-based methods \cite{KleinMAP}. These methods mainly differ in how efficient they are (i.e. how many evaluations of the deterministic model are needed to obtain an accurate result), how much information about the desired probability distribution they deliver, and how many assumptions on the model they make.

Many of these methods simply treat the deterministic model as a map $F : \mathbb{R}^n \rightarrow \mathbb{R}^m$, taking an $n$-dimensional parameter to an $m$-dimensional model output. In addition to evaluations $F(\theta)$, some methods require derivatives in the form of Jacobian actions, gradients, or Hessian actions.

This common mathematical \textit{interface} between method and model means that applying any such method to any model providing the aforementioned operations is, in theory, trivial. However, advanced \gls{UQ} methods and state-of-the-art numerical models often necessitate complex software packages. Examples for \gls{UQ} software might include PyMC \cite{pyMC3}, \gls{sgmk} \cite{sparse_grids_matlab_kit} or \gls{MUQ} \cite{MUQ}. On the model side, established packages include DUNE \cite{DUNE}, ExaHyPE \cite{ExaHyPE} or deal.II \cite{dealII94}.\footnote{We list some packages currently in use with UM-Bridge; this is not intended as a representative or comprehensive list of available packages.} Directly coupling \gls{UQ} software to numerical models is often disproportionately challenging for a number of reasons:
\begin{itemize}
    \item \textbf{Software compatibility} may be time consuming or impossible to achieve due to the different programming languages or frameworks being used (e.g. R or MATLAB for \gls{UQ}, C++ or Fortran for \gls{PDE} solvers), different operating systems supported, incompatible build systems, dependency conflicts etc,
    \item \textbf{Separation of concerns} is nearly impossible since directly coupling two code-bases requires experts from both sides to work together,
    \item \textbf{\gls{HPC}} scale applications are hard to achieve when the resulting software complexity makes installation on \gls{HPC} systems a challenge or when incompatible parallelization concepts are used (e.g. thread parallel software interacting with MPI \cite{MPI} based codes).
\end{itemize}

The \gls{UMBridge} \cite{UMBridge} addresses these issues by providing a network based abstraction of models. Inspired by microservice architectures, method and model code are now two independent applications. Requests for model evaluations and their results are passed between the two through an \gls{HTTP} based protocol, essentially implementing remote procedure calls for mathematical functions and their derivatives. Native integrations for a number of languages and frameworks make calling a model as easy as a regular function call in the respective language. Since it is network-based, any \gls{UMBridge} client may connect to any \gls{UMBridge} supporting model regardless of the respective programming languages or frameworks.

\gls{UMBridge} models are called via network, so models can run locally, inside a container (e.g. docker  \cite{docker} or Apptainer \cite{singularity}), or even on a remote cluster. Containerization replaces model software installation by a simple download, and allows models to run regardless of operating system, software environment etc. This leads to strong separation of concerns, since the model expert may exclusively take care of building the model container, and the \gls{UQ} expert in turn, may treat the model software as a black box.

Finally, containerization opens the option to immediately, and without modification, run containerized \gls{UMBridge} models on modern cloud environments such as GCP or AWS. To that end, the \gls{UMBridge} project provides a reference configuration for the widely used kubernetes orchestrator, reducing setup effort to copying the reference configuration and choosing a custom model container image to run. Parallel model instances for \gls{HPC} scale performance are fully handled by the reference setups, while even running model instances each spanning multiple hardware nodes via \gls{MPI} can be achieved with only minor modifications to the model image.  Most importantly, distribution of work is fully handled inside the cluster, allowing \gls{UQ} software to simply request model evaluations in parallel without having to implement load balancing mechanisms.

We begin with an overview of mathematical background and \gls{UMBridge} architecture in \cref{sec:umbridge}, accompanied by basic examples.
The main contributions of this paper are the kubernetes setup for scaling any UM-Bridge model from single machines to large-scale cloud clusters (\cref{sec:kubernetes}), as well as a number of advanced \gls{UQ} applications demonstrating the effectiveness of UM-Bridge in \cref{sec:applications}. In particular, UM-Bridge enables existing prototype-grade \gls{UQ} applications to transparently control clusters of thousands of processor cores (\cref{sec:application_mlda}).

%% file: umbridge.tex
\section{UM-Bridge}
\label{sec:umbridge}


\subsection{Mathematical model interface}
\label{sec:math_interface}


Many \gls{UQ} methods consider a deterministic model as a map $F : \mathbb{R}^n \rightarrow \mathbb{R}^m$. The model map takes a parameter vector $\theta$ of dimensions $n$ to some model output, a vector of dimension $m$.
The model is often assumed to provide some of the following mathematical operations:
\begin{itemize}
    \item Evaluation $F(\theta)$,
    \item gradient $v^\top J_F(\theta)$,
    \item Jacobian action $J_F(\theta) v$, and
    \item Hessian action.
\end{itemize}


Such a \gls{UQ} method could therefore - in principle - be applied to any model supporting the appropriate operations.

Forward \gls{UQ} problems are defined via a random vector 
$\theta$ following some distribution. The goal is to find properties of the distribution of $F(\theta)$,
such as moments, quantiles, \gls{pdf} or cumulative distribution function.
Many forward \gls{UQ} methods (e.g. \gls{MC} \cite{MC} or stochastic collocation \cite{SC_Survey})
only require model evaluations $\{F(\theta_i)\}_{i=1}^N$ at a finite number of points.

For inverse \gls{UQ} problems, the deterministic model $F$ becomes part of some form of distance measure between model prediction and observed data. Specifically, Bayesian inference defines the probability of a parameter $\theta$ given observed data $y$ in terms of a prior and a likelihood distribution:
\[ \pi_{post}(\theta | y) \propto \pi_{prior}(\theta) \pi_{likelihood}(y | \theta), \]

The former represents prior information about the desired parameter distribution, while the likelihood depends on the distance between model prediction and data.

The goal of inverse UQ is then to generate samples of $\theta$ according to $\pi_{post}$, for which different methods can 
be employed (Metropolis-Hastings MCMC \cite{MHMCMC}, Hamiltonian MCMC \cite{HamiltonianMCMC}, NUTS \cite{NUTS}, etc.), which also differ in requirements on 
the model: some only require model evaluations, others require the gradient or even the Hessian.

It should be noted that multilevel, multiindex or multifidelity methods (e.g. \cite{MLMC,MLMCMC,Lykkegaard_MLDA,SchwabMLMCMC,MultilevelDILI,lykkegaard_multilevel_2023,teckentrup.etal:MLSC,MIMCMC,piazzola.eal:ferry-paper,sandia:coupled}) like the one in \cref{sec:application_mlda} operate on an entire hierarchy of models. Less accurate but cheap to compute models are then used to accelerate the \gls{UQ} method, leading to considerable gains in efficiency both in forward and inverse \gls{UQ} problems. While this necessitates multiple models, each individual model still typically fits in the framework described above.

\subsection{UM-Bridge interface}

\begin{figure}
\centering

%
\includegraphics[width=0.4\columnwidth]{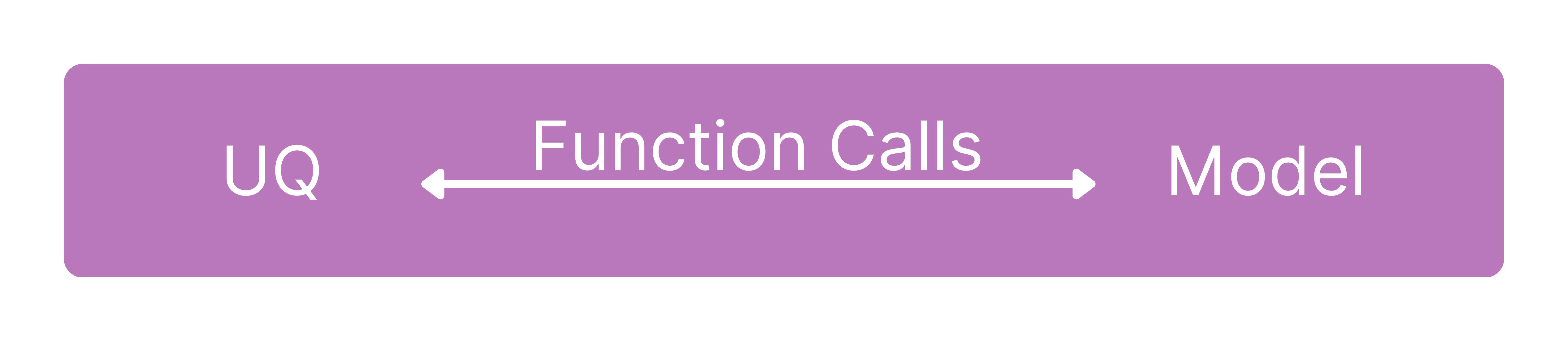}
%
%
%
\includegraphics[width=0.4\columnwidth]{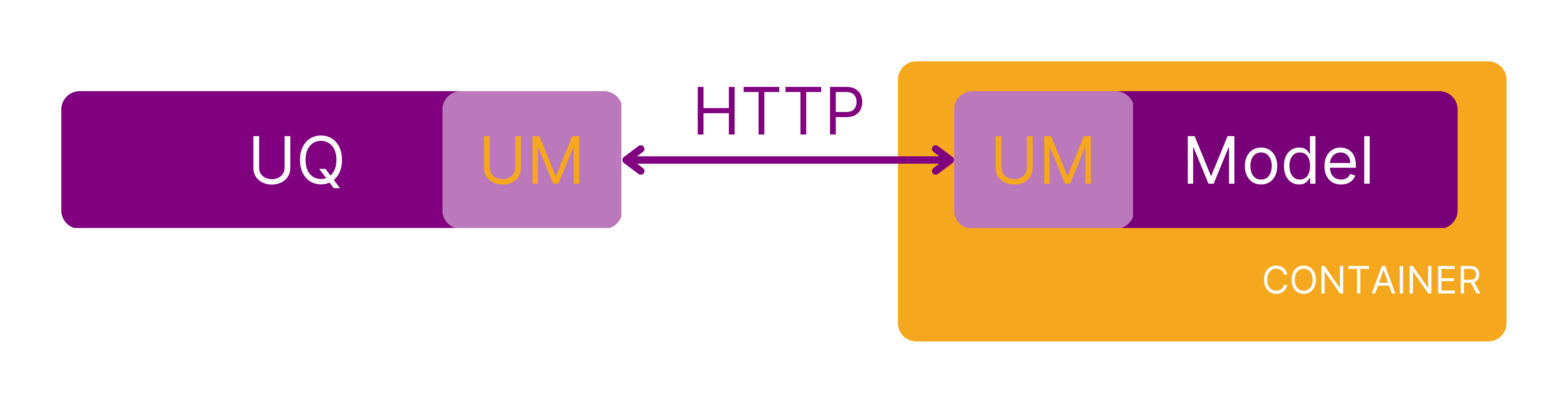}

\caption{Left: Monolithic coupling of \gls{UQ} and model software in a single application. Right: UM-Bridge providing a universal interface between \gls{UQ} and model. Optionally, an UM-Bridge model may be containerized.}
\label{fig:microservice}
\end{figure}

UM-Bridge mirrors this mathematical \textit{interface} from \cref{sec:math_interface} as an equally universal software interface. Instead of combining \gls{UQ} and model software in a single complex application, UM-Bridge is an \gls{HTTP} based protocol linking model and \gls{UQ} software through network communication (see \cref{fig:microservice}). The model software is now acting as a server accepting model evaluation requests from \gls{UQ} clients.

In contrast to a monolithic approach, this microservice-inspired architecture offers a number of opportunities:

\begin{itemize}
  \item Level 1: Methods and models can easily be coupled across arbitrary languages and frameworks. Easy-to-use integrations for a number of languages are available, handling all network communication behind the scenes (\cref{tab:lang_support}).
  \item Level 2: Through UM-Bridge, models can easily be containerized \cite{singularity}. We obtain strong separation of concerns between model and method experts, accelerating development. Further, reproducibility of models becomes possible, and a \gls{UQ} benchmark library based on UM-Bridge is currently under development as a community effort.
  \item Level 3: Scaling small-scale \gls{UQ} applications to HPC-scale cloud systems becomes possible. Prototype-grade, thread-parallel \gls{UQ} method codes can control entire clusters running computationally expensive models.
\end{itemize}

\subsection{Language and framework support}

\begin{table}
  \center
\begin{center}
  \begin{tabular}{||c c c||}
   \hline
   Language / Framework & Client & Server \\ [0.5ex]
    \hline\hline
    Python & \cmark & \cmark \\
    C++ & \cmark & \cmark \\
    MATLAB & \cmark & \xmark \\
    R & \cmark & \xmark \\
    Julia & planned & \xmark \\
    \hline\hline
    MUQ \cite{MUQ} & \cmark & \cmark \\
    PyMC \cite{pyMC3} & \cmark & \xmark \\
    QMCPy \cite{QMCPy} & \cmark & \xmark \\
    \gls{sgmk} \cite{sparse_grids_matlab_kit} & \cmark & \xmark \\
    tinyDA \cite{tinyDA} & \cmark & \xmark \\
   \hline
  \end{tabular}
\end{center}
  \caption{Client and server support in various languages and frameworks.}
  \label{tab:lang_support}
\end{table}

Generally, the UM-Bridge protocol can be implemented in any language or framework supporting HTTP requests and JSON. A number of native integrations for various languages and frameworks is provided (see \Cref{tab:lang_support}). These integrations take care of network communication behind the scenes, and make UM-Bridge models available either as simple function calls or as fully integrated models in the respective framework's model structure.

\subsection{Basic examples}
\label{sec:howto}

This section briefly walks through the workflow of setting up an UM-Bridge client and server for a dummy application. While intentionally basic, the same approach underlies the applications in \Cref{sec:applications}.

\subsubsection{Clients}\label{sec:clients}

A basic UM-Bridge client can be created with very little effort. For example, in Python an UM-Bridge model is called as follows via the \texttt{umbridge} module:

\begin{lstlisting}[language=Python,breaklines=true,postbreak=\mbox{$\hookrightarrow$\space}]
url = "http://localhost:4242"
model = umbridge.HTTPModel(url, "forward")

print(model([[0.0, 10.0]]))
\end{lstlisting}

All that was needed is the url at which the model server is running, here \verb=http://localhost:4242=, the requested model, here \verb=forward=, to evaluate $F$ at, and a parameter $\theta$. The actual evaluation is then a call to the model with a given parameter $\theta=(0,10)$, which returns $F(\theta)$. Optionally, we can pass arbitrary configuration options to the model:

\begin{lstlisting}[language=Python]
print(model([[0.0, 10.0]], {"level": 0}))
\end{lstlisting}

Similar calls allow access to gradient, Jacobian action and Hessian action, as long as the model supports them.

The model client will follow a similar format in other supported languages. For example, the exact same model can be called from C++ via:

\begin{lstlisting}[language=C++]
std::string url = "http://localhost:4242";
umbridge::HTTPModel model(url, "forward");

std::vector<std::vector<double>> outputs
= model.Evaluate({{0.0, 10.0}});
\end{lstlisting}

\subsubsection{Servers}
Model servers conversely require implementation of the model map $F$. Model name, input and output dimensions $n$ and $m$ and model evaluation itself must be provided. This is a minimal working example in Python, multiplying the single input value by two:

\begin{lstlisting}[language=Python]
class TestModel(umbridge.Model):

  def __init__(self):
    super().__init__("forward")

  def get_input_sizes(self, config):
    return [1] # Input dimensions

  def get_output_sizes(self, config):
    return [1] # Output dimensions

  def __call__(self, parameters, config):
    output = parameters[0][0] * 2
    return [[output]]

  def supports_evaluate(self):
    return True

umbridge.serve_models([TestModel()], 4242)
\end{lstlisting}

The model also specifies \verb=supports_evaluate=, telling UM-Bridge that model evaluations are supported. Likewise, gradients, Jacobian action or Hessian action could be implemented and marked as supported.

Similarly, UM-Bridge servers can be implemented in C++ as well.

\subsubsection{Containers}

Many numerical models consist of a complex software stack and possibly associated data sets, such that each installation requires considerable effort and expert knowledge. Since UM-Bridge models are accessed through network however, they can easily be containerized. Such a container can then be launched on any local machine or remote cluster without installation effort, and the model accessed through UM-Bridge as before. Containerized models can therefore easily be shared among collaborators, achieving separation of concerns between model and \gls{UQ} experts.

Defining such a container comes down to instructions for installing dependencies, compiling if needed, and finally running the desired UM-Bridge server code. For example, a docker container wrapping the example server above can be defined in the following simple Dockerfile:

\begin{lstlisting}
FROM ubuntu:latest

COPY minimal-server.py /

RUN apt update && \
    DEBIAN_FRONTEND="noninteractive" apt install -y python3-pip && \
    pip install umbridge

CMD python3 minimal-server.py
\end{lstlisting}

A special use case are model codes that are not suited to be integrated in another application, not even an UM-Bridge server. In this case the UM-Bridge server can be a small wrapper application that, on every evaluation request, launches the actual model software with appropriate configuration. Such a setup could be error-prone to replicate on another machine (and especially clusters), but is perfectly reliable inside a container.


%% file: scalability.tex
\section{Scaling up UM-Bridge applications}
\label{sec:kubernetes}

\subsection{UM-Bridge on kubernetes clusters}

We use kubernetes in order to transparently scale up UQ applications on clusters, enabling HPC-scale UQ applications on even prototype-grade UQ software and any UM-Bridge supporting model. The reasoning behind using kubernetes is that

\begin{itemize}
    \item existing model containers can be run without modification,
    \item the entire setup can be specified in configuration files and is easily reproducible,
    \item and it can be run on systems ranging from single servers to large-scale cloud systems.
\end{itemize}

As part of the UM-Bridge project, we provide a kubernetes configuration that users can easily adapt to their own models. Its architecture and use is described below.

\subsubsection{Architecture}

\begin{figure*}
\centering

\includegraphics[width=0.9\textwidth]{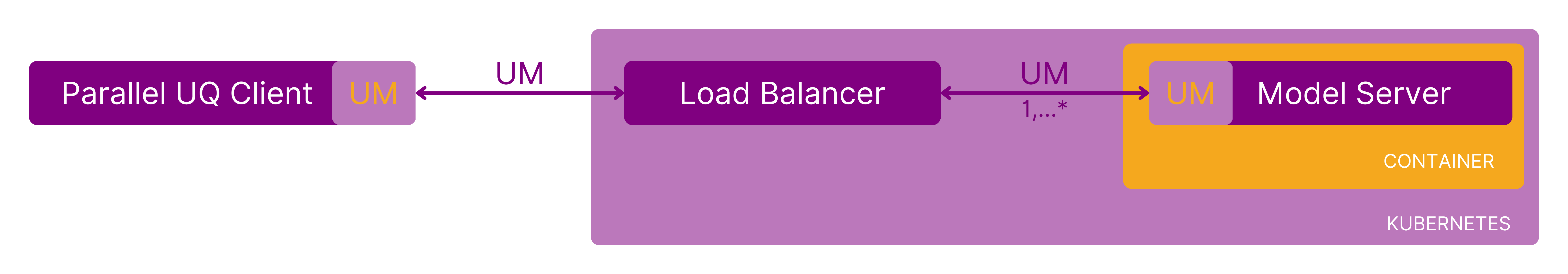}
%
%
%
%
%
%
\caption{Kubernetes setup for parallel instances of single-container models.}
\label{fig:kubernetes_sequential_model}
\end{figure*}

On kubernetes, we spin up an arbitrary number of model containers that can be accessed from outside the cluster through a load balancer (see \Cref{fig:kubernetes_sequential_model}). The load balancer distributes model evaluation requests across all model instances in the cluster.

We employ HAProxy as load balancer. It is originally intended for web services processing large amounts of requests per backend container. However, we set it up such that only a single evaluation request is ever sent to a model server at once, since we assume running multiple numerical model evaluations concurrently on a single machine would lead to performance degradation.

As the interface is identical, a sequential UQ software is completely oblivious to the model being executed on a cluster instead of a local machine. The UQ software may now send multiple parallel evaluation requests to the cluster. Relatively simple thread parallelism is therefore sufficient on the UQ side. Since the model evaluation is typically costly compared to the UQ algorithm itself, a UQ package spawning hundreds of threads on a laptop controlling an entire cluster is perfectly viable. Since the UQ software is not involved in distributing evaluations across the cluster, the existing parallelism in numerous UQ packages works out of the box with this setup. The \gls{sgmk}, MUQ, QMCPy, and PyMC have been tested successfully. Some of these applications are shown in \Cref{sec:applications}.

\subsubsection{Usage}

Kubernetes can be fully controlled through configuration files. The UM-Bridge project provides a reference configuration implementing the architecture above. It can be applied as is on any kubernetes cluster by cloning the UM-Bridge Git repository and executing

\begin{lstlisting}
kubectl apply -f FILENAME
\end{lstlisting}

on the provided configuration files. This procedure is described in more detail in the UM-Bridge documentation, and typically takes only a few minutes.

The only model specific changes needed for a custom application are docker image name, number of instances (\texttt{replicas}) and resource requirements in the provided \texttt{model.yaml} configuration. Below is an example of an adapted model configuration, specifically the one used in the sparse grids application in \Cref{sec:application_sparse_grids}.

\begin{lstlisting}
apiVersion: apps/v1
kind: Deployment
metadata:
  name: model-deployment
spec:
  replicas: 48
  template:
    metadata:
      labels:
        app: model
    spec:
      containers:
      - name: model
        image: linusseelinger/model-l2-sea:latest
        env:
        - name: OMP_NUM_THREADS
          value: "1"
        resources:
          requests:
            cpu: 1
            memory: 1Gi
          limits:
            cpu: 1
            memory: 1Gi
\end{lstlisting}

\subsection{Models parallelized across nodes}

\begin{figure*}
\centering

\includegraphics[width=0.9\textwidth]{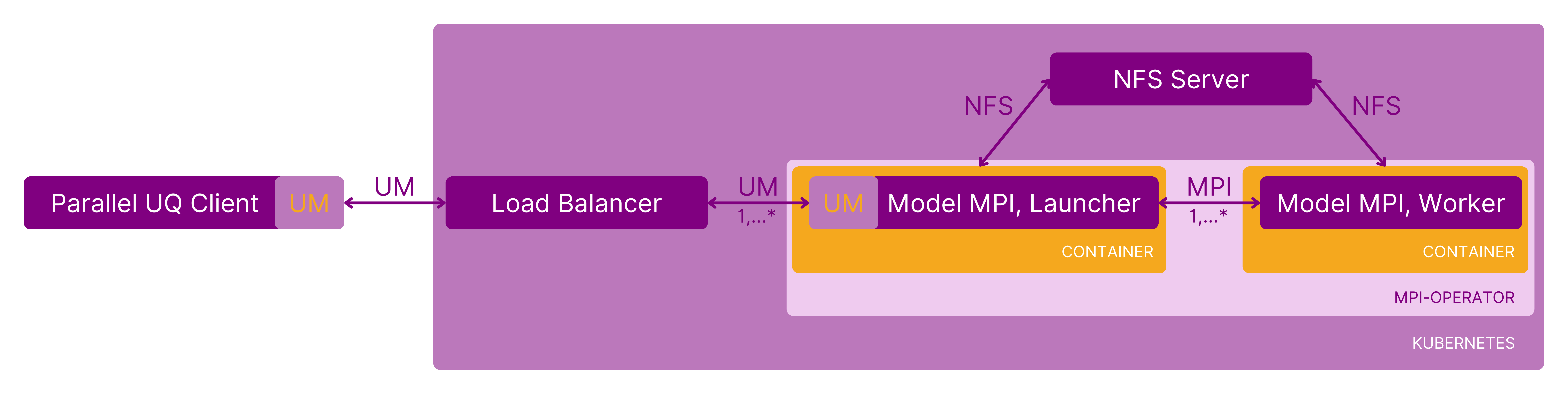}
\caption{Kubernetes setup for MPI-parallel models with shared filesystem via NFS.}
\label{fig:kubernetes_mpi_parallel_model}
\end{figure*}

A container is inherently confined to running on a single physical machine. However, many numerical models can (or even have to) be parallelized across physical machines, typically via MPI \cite{MPI}.
In order to support these, we provide an extended kubernetes configuration allowing for models to be MPI-parallel across containers (see \Cref{fig:kubernetes_mpi_parallel_model}).

The main difference is that, in order to span multiple physical machines, models are separated into a launcher and multiple workers containers. They may now use regular MPI operations to communicate. Internally, this is achieved using on kubeflow's \texttt{mpi-operator}, which is currently limited to OpenMPI and Intel MPI. As another restriction, model containers now have to inherit from one of the base images provided by \texttt{mpi-operator}, or themselves perform some simple preparatory operations that these image already provide.

In addition to MPI parallelization of each model instance, multiple independent models may be launched, achieving the same model parallelism as before. Again, the load balancer distributes incoming requests across all available model instances.

Evaluation requests from UM-Bridge arrive at the launcher container, which should therefore spawn the UM-Bridge server. Workers in turn can be identical to the launcher, but do not execute the model. Instead, they are available as potential MPI ranks when the launcher issues \texttt{mpirun} commands.

Since some models assume a shared file system between workers (and possibly the launcher as well), an optional NFS server is integrated in the kubernetes configuration. Its shared file system can be mounted by launcher and workers and operated on akin to an HPC system.

As before, this kubernetes configuration can immediately be applied on any kubernetes cluster and is fully transparent to the UQ software.

\subsection{Synthetic scalability test}

This test demonstrates weak scalability in the sense of increasing both model instances on a remote cluster and evaluations requested from it. The overhead to be expected is mainly from network communication, including the load balancer.

We placed a kubernetes cluster on the \gls{GKE} with the setup of \Cref{sec:kubernetes} under a representative synthetic load. As a representative model, we choose the L2-Sea model from the UM-Bridge benchmark library. Each model instance is configured to require one processor core, and takes about $2.5s$ per evaluation. For parallel requests, we set up a parallel \gls{MC} sampler implemented in \gls{MUQ}, with the number of threads matching the number of models on the cluster. It then requests model evaluations from the cluster. We artificially force \gls{MUQ} to always sample the same parameter in order to avoid significant run time fluctuations in the model itself. We disable \gls{SMT} on \gls{GKE}, since it barely offers any overall performance benefit in our application, but leads to strongly varying model run times.

\begin{figure}
\centering
\resizebox{0.8\columnwidth}{!}{
\begin{tikzpicture}
	\begin{axis}[
		xlabel=Number of processes,
		ylabel=Run time {[}s{]},
		grid=major,
		xmode=log,
		ymin=0,
        height=4cm,
        width=\linewidth
	]

	\addplot coordinates {
		(512,49.429) 
		(256,53.398) 
		(128,50.534) 
		(64,50.058) 
		(32,51.400) 
		(16,47.233) 
		(8,47.147) 
		(4,46.930) 
		(2,46.956) 
		(1,46.901) 
	};
	\end{axis}
\end{tikzpicture}
}
\caption{Run time of synthetic scalability test of the kubernetes setup in \Cref{sec:kubernetes} on \gls{GKE} for various numbers of parallel model instances.}
\label{fig:cloud_scalabillity}
\end{figure}

The results in \Cref{fig:cloud_scalabillity} show near perfect weak scaling, indicating low parallel overhead. Some fluctuations are observed for higher process counts, which we believe comes mostly from non-exclusive access to physical nodes in the cloud system.

We did not observe a bottleneck in testing, and assume this setup could be scaled further. In addition, the model itself could be parallelized. For example with a model parallelized across 20 cores, we can expect similar results to the ones presented here on a 10,000 core cluster.

%% file: applications.tex
\section{Applications}
\label{sec:applications}

The following sections present \gls{UQ} applications that UM-Bridge enabled through
\begin{itemize}
    \item straightforward linking of \gls{UQ} and model codes despite potential incompatibilities,
    \item containerized models being shared among collaborators, greatly improving separation of concerns and accelerating development,
    \item scaling applications from small scale local testing all the way to large cloud clusters,
    \item and allowing existing \gls{UQ} packages with various simple parallelization methods to control these clusters.
\end{itemize}


\Cref{sec:application_sparse_grids} shows how an existing Matlab code for sparse grid \gls{UQ} methods can, without modification, transparently control a number of model instances running on a cloud system. \Cref{sec:application_qmc} demonstrates parallel runs of a highly complex numerical solver in a \gls{QMC} application, highlighting the need for portable models and separation of concerns between model and \gls{UQ} experts.
Finally, \Cref{sec:application_mlda} demonstrates a complex \gls{UQ} application combining both fast approximations running on a workstation and costly parallel model evaluations offloaded to a cloud cluster of 2800 physical processor cores.

\subsection{Accelerating a naval engineering application}
\label{sec:application_sparse_grids}

\begin{figure*}
    \centering
    \includegraphics[trim={200 60 150 40},clip,width=\linewidth]{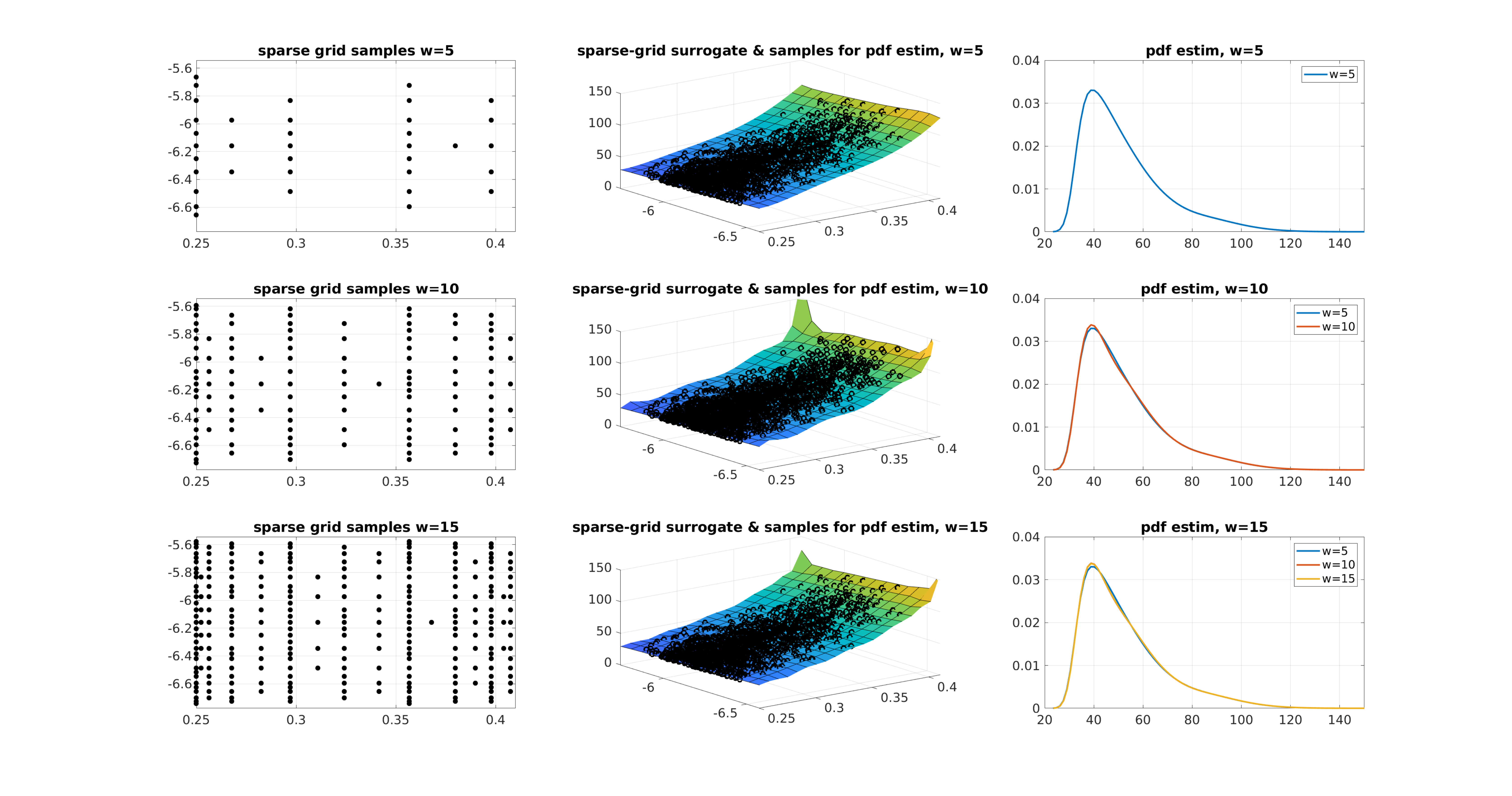}
    \caption{Results for the UQ workflow on the L2-Sea model for three 
    sparse-grid levels $w$ (one row for each $w$). Left column: 
    sparse grids covering the space of possible values of $F,D$; central column: sparse-grid surrogate models and their evaluations at the random sample points used for computing the \gls{pdf} of $R_T$; right column: resulting \gls{pdf} of $R_T$ for all $w$ up to the current one.}
    \label{fig:results_Lorenzo}
\end{figure*}

\subsubsection{Problem description}
In this section we focus on a forward UQ application in the context of 
naval engineering. The goal is to compute the \gls{pdf}
of the resistance to advancement $R_T$ of a boat (naval equivalent of the drag force for airplanes).
The boat is advancing in calm water, under uncertain
Froude number $F$ (a dimensional number proportional to the navigation speed)
and draft $D$ (immersed portion of the hull, directly proportional to the payload),
i.e., $\theta = [F,D]$. The computation of $R_T$ for fixed $F,D$,
i.e. the evaluation of the response function $R_T = R_T(F,D)$,
is performed by the L2-Sea model \cite{serani:l2sea}. This model is written in Fortran, but has an UM-Bridge wrapper and is available as a container from the UM-Bridge benchmark library

In order to showcase the ease of parallelizing an existing application by using UM-Bridge, we will show the required command and code snippets in this section. The applications in subsequent sections follow the same approach. This model can be run locally via the following docker command: 
\begin{lstlisting}
docker run -it -p 4242:4242 linusseelinger/model-l2-sea
\end{lstlisting}
 
The uncertain parameters are modeled as follows: 
Froude is a triangular random variable with support over $[F_a,F_b]=[0.25, 0.41]$, 
$F \sim Triang(F_a;F_b)$ 
while draft is a beta random variable with support over $[D_a,D_b]=[-6.776, -5.544]$
and shape parameters $\alpha=10,\beta=10$, $D \sim Beta(D_a,D_b,\alpha,\beta)$
\footnote{several parametrizations of Beta random
variables are possible; the form of the \gls{pdf} considered here reads
$\rho_{\mathrm{Beta}}(x) = \frac{\Gamma(\alpha+\beta+2)}{\Gamma(\alpha+1)\Gamma(\beta+1)} (D_b-D_a)^{\alpha+\beta+1}(x-D_a)^\alpha(D_b-x)^\beta$}.

\subsubsection{UQ workflow}
To compute the probability density function of $R_T$ we proceed in two steps: 
\begin{enumerate}
    \item we create a surrogate model for the response function $R_T = R_T(F,D)$, 
    i.e. an approximation of the actual function based on a limited 
    number of (judiciously chosen) evaluations of $R_T = R_T(F,D)$. 
    
    \item we generate a large sample of values of $F,D$ according to their \gls{pdf},
    we evaluate the surrogate model for each element of the sample
    (which is considerably cheaper than evaluating the full model $R_T(F,D)$, but yields only approximate results), and then use the corresponding values of $R_T$ to compute an 
    approximation of the \gls{pdf} of $R_T$ by e.g. kernel density algorithms \cite{rosenblatt:kde}.
\end{enumerate}

To build the surrogate model, we consider the sparse-grids method, and in particular the implementation provided by \gls{sgmk}. This method
requires sampling the space of feasible values of $F,D$ with a 
structured non-cartesian strategy, depending on the \gls{pdf} of the two uncertain parameters, 
see the left column of \Cref{fig:results_Lorenzo}. The surrogate model is constructed as a sum of certain interpolating polynomials, each based on a subset of samples. The surrogate models are reported in the central column \Cref{fig:results_Lorenzo}. The overshoots in the corners of the domain happen in regions of the parameter space of \textit{zero probability}; neither full model evaluations nor samples to approximate the \gls{pdf} of $R_T$ are placed there.

\gls{sgmk} provides ready-to-use functions to generate sparse grids according to several types of 
random variables (uniform, normal, exponential, beta, gamma, triangular);  
once the grid is created, it is a simple matter of looping through its points
and calling the L2-Sea solver for each one using the point coordinates as values for the inputs $F,D$. 
In \gls{sgmk} this is as easy as the following snippet: 
\begin{lstlisting}[language=Matlab,keywords={}]
% uri of L2-Sea solver
  uri = 'http://104.199.68.148'; 
  model = HTTPModel(uri,'forward');
% setting up L2-Sea configuration
  config=struct('fidelity',3,'sinkoff','y','trimoff','y')
% L2-Sea takes 16 inputs but we use only the first two, 
% we need to fix the others
  inputs = @(y) [y' zeros(1,14)];
% wrap UM-Bridge call for compatibility with SGMK
  f = @(y) model.evaluate(inputs(y),config);

% functions to generate nodes for F and D
  x1_a = 0.25; x1_b =0.41;
  knots_Froude = @(n) knots_triangular_leja(n,x1_a,x1_b);
  x2_a=-6.776; x2_b=-5.544; alpha=10; beta=10;
  knots_Draft = @(n) knots_beta_leja(n,alpha,beta,x2_a,x2_b,'sym_line','on_file');

% build sparse grid (last argument omitted for brevity)
  N=2; w=5;
  S = smolyak_grid(N,w,{knots_Froude,knots_Draft},...);
  Sr = reduce_sparse_grid(S);

% call L2-Sea on each point, possibly use Matlab's parfor 
% to allocate different points to different workers 
% (optional arguments omitted for brevity)
  f_values = evaluate_on_sparse_grid(f,Sr,...);
\end{lstlisting}

Once obtained \lstinline{f_values}, we can move the second step of the procedure 
outlined above. We generate a random sample of values of $F,D$
according to their \gls{pdf} (e.g. by rejection sampling \cite{casella2004generalized}), and  
evaluate the surrogate model on such values with a straightforward call
to a \gls{sgmk} function: 
\begin{lstlisting}
surrogate_evals = interpolate_on_sparse_grid(S,Sr,f_values,random_sample);
\end{lstlisting}
These evaluations are shown as black dots in the central column of \Cref{fig:results_Lorenzo}.
These surrogate model evaluations are finally input to the standard ksdensity estimator provided by Matlab to compute an approximation of the \gls{pdf} of $R_T$
\begin{lstlisting}
[ksd_pdf,ksd_points] = ksdensity(surrogate_evals,'support','positive','Bandwidth',0.1);
\end{lstlisting}
Results are reported in the right column of \Cref{fig:results_Lorenzo}, and as expected, 
the estimated \gls{pdf} stabilizes as we add more points to the 
sparse grid and the surrogate model becomes more reliable. The three rows are
obtained repeating the snippets above three times, increasing the so-called 
\textit{sparse grid level} \lstinline{w}, which is 
an integer value  that controls how many points are to 
be used in the sparse grid procedure, specifically \lstinline{w=5,10,15}, corresponding
to $36,121,256$ points. Note that the three sparse grids produced are 
nested, i.e., they are a sub-set of one another, so that in total only $256$ calls to L2-Sea are needed.
 \gls{sgmk} is able to take advantage of this and only evaluate at the 
\textit{new} sparse-grid points in each grid level.

\subsubsection{Computational aspects}
In order to allow parallel model evaluations, 48 instances of the model were run on \texttt{c2d-high-cpu} nodes of \gls{GKE} using the kubernetes configuration in \Cref{sec:kubernetes}. For the test case presented here, one evaluation of the L2-Sea model takes about 30-35s on a single physical CPU core. Note that we disable \gls{SMT} on \gls{GKE}, since it offers nearly no efficiency gains in our applications, but leads to significant variations in model run times. Nevertheless, minor variations are observed depending on the specific values of the parameters $F,D$.

On the \gls{UQ} side, \gls{sgmk} was run on a regular laptop, connecting to the cluster. Due to UM-Bridge, the \gls{sgmk} Matlab code can directly be coupled to the Fortran model code. Further, no modification to \gls{sgmk} is necessary in order to issue parallel evaluation requests: The function \lstinline{f_values = evaluate_on_sparse_grid(f,Sr,...)} in \gls{sgmk}
is internally already using Matlab's \lstinline{parfor} in order to loop over parameters to be evaluated. By opening a parallel session in Matlab with $48$ workers 
(via \lstinline{parpool} command), the very same code executes the requests in parallel, and the cluster transparently distributes requests across model instances.

By using the cluster we have obtained a total run time of around $290$ seconds instead of the initial
$30 \times 256 = 7680$ seconds (i.e., 2h10m), achieving a speedup of around $26.5$.
Three factors affect the speedup: The evaluations are split in three batches, none of the batches having cardinality equal to a multiple of the $48$ parallel model instances; the CPU-time for evaluating the L2-Sea model at different values of $F,D$ is not identical; and finally a minor overhead in the \gls{UQ} method itself.

Apart from the limited number of evaluations needed for the specific application, the main limit in scalability we observe here is Matlab's RAM consumption per worker with approximately 500 Mb allocated per worker, regardless of the actual workload. This could be improved by e.g. sending batches of requests per worker, but not without modifying the existing \gls{sgmk} code.



\subsection{Efficient investigation of material defects in composite aero-structures}
\label{sec:application_qmc}

\subsubsection{Problem description}

In this section we examine the effect of random material defects on structural aerospace components made out of composite laminated material. These may span several meters in length, but are still governed by a mechanical response to processes on lower length scales (sub-millimetres). Such a multi-scale material displays an intricate link between meso-scale (ply scale) and macro-scale behaviour (global geometric features) that cannot be captured with a model describing only the macro-scale response. This link between material scales is even stronger as abnormalities or defects, 
such as localised delamination between plies, may occur during the fabrication process.

In order to avoid a large number of very costly full scale model runs, a multiscale spectral generalized finite element method (MS-GFEM) \cite{Benezech:2023} is used.
The approach constructs a coarse approximation space (or reduced order model) of the full-scale underlying composite problem by solving local generalized eigenproblems in an A-harmonic subspace. 
This multiscale method has a major advantage in the context of localised defects: Assuming each subsequent problem is identical except for changes localized to a few subdomains (such as a localized defect in an aerospace part), the approximation space can remain identical in all other subdomains and largely be reused across runs. An offline-online framework is proposed: Offline, the MS-GFEM is applied to a pristine model, and online, only the eigenproblems on subdomains intersecting local defects are recomputed to update the approximation space. Online recomputations can be carried out by a single processor, freeing resources for multiple parallel simulation runs.

\begin{figure}[t!]
    \centering
    \includegraphics[width=0.49\linewidth]{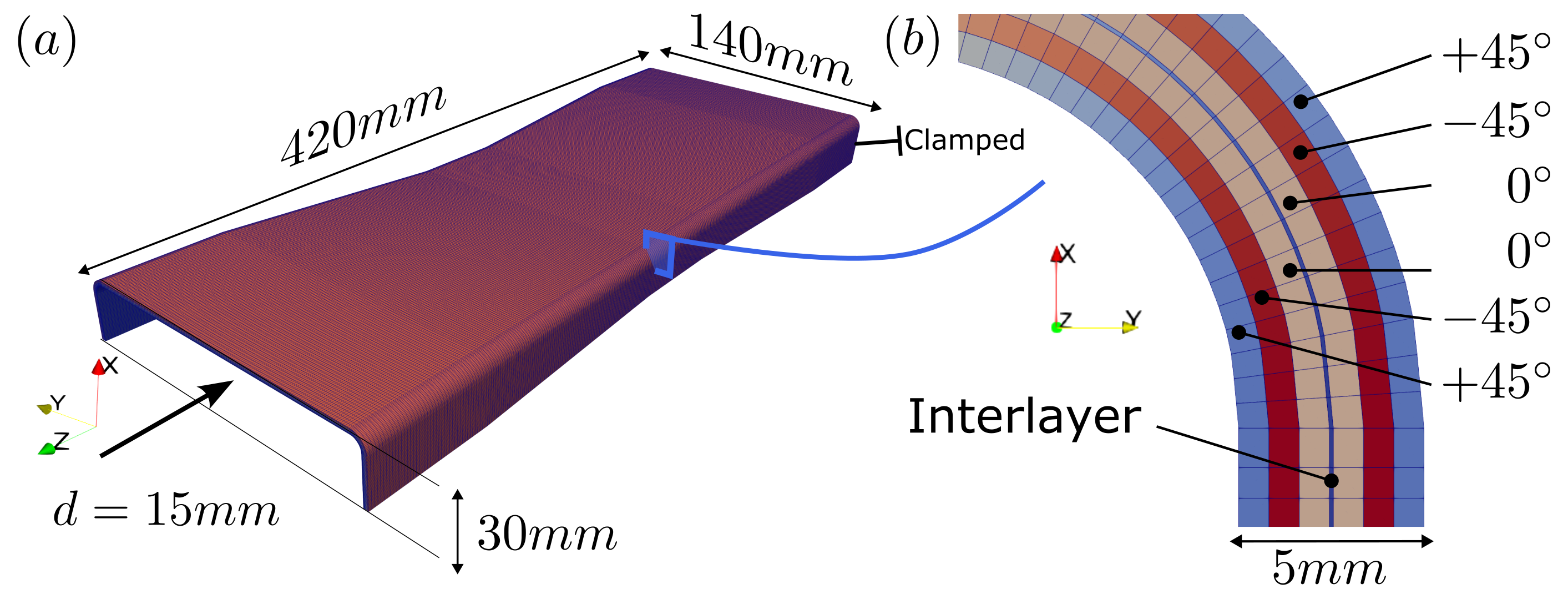}
    \includegraphics[width=.31\linewidth]{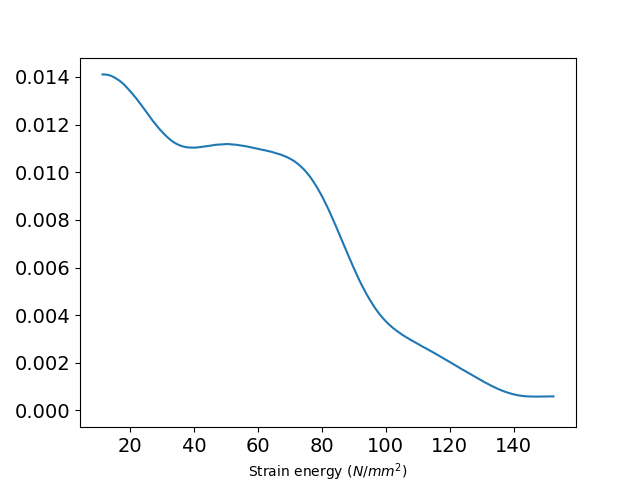}
    \includegraphics[width=.17\linewidth]{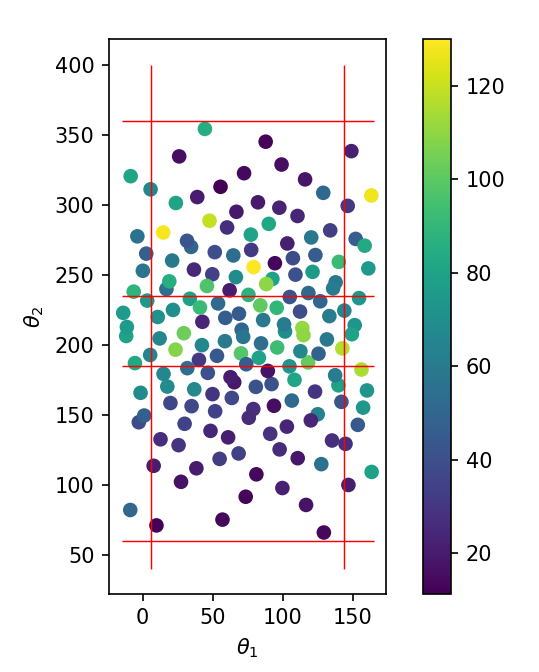}
        \caption{Left: Composite laminated C-spar model: Geometric description at macro-scale (a) and through thickness discretization at meso-scale (b).
   Center: Distribution of strain energy resulting from randomly distributed defect. Right: Spatial distribution of samples (defect size is not indicated).}
    \label{fig:composites_results}
\end{figure}

Specifically, the behavior of an aerospace component under compression has been simulated (namely a laminated composite C-shaped spar, see~\Cref{fig:composites_results}(a)) . 
The laminate consists of a six-layer stack of uni-directional composites made up of carbon fibres embedded in resin, which are oriented following the stacking sequence described in~\Cref{fig:composites_results}(b).
To model the delamination defect, a resin interlayer between the two central composite plies is introduced. The delamination defect is represented by a local reduction of the elastic property of the resin interlayer. To assess the severity of the defect, the strain energy is computed. 
The full scale model consists of two million degrees of freedom. The MS-GFEM approximation reduces it to $32,721$, which corresponds to a model order reduction factor of $\approx 58$. 

The method is implemented in C++ and integrated in the \gls{DUNE} \cite{DUNE}.

\subsubsection{UQ workflow}

We solve a forward \gls{UQ} problem by assuming a random defect parameter $$\theta \sim \mathcal{N}((77.5, 210, 10), \text{diag}(8000, 4800, 2))$$ with Gaussian distribution, which we cut off at the domain boundary. The first two components $\theta_1$ and $\theta_2$ refer to the defect position along width and length of the part, while $\theta_3$ is the diameter of the defective region. Each value is given in mm. The goal is to obtain the resulting distribution of a failure criterion, namely maximum strain energy.


The parameter $\theta$ is sampled via \gls{QMC}, specifically using QMCPy's \texttt{CubQMCSobolG} implementation of Sobol' cubature. From 256 samples, we obtain the failure criterion distribution in \Cref{fig:composites_results}.

\subsubsection{Computational aspects}

The numerical experiment has been conducted by running the model container on a 36-core machine, which was outfitted with a simple k3s installation in order to support the kubernetes setup of \Cref{sec:kubernetes}. QMCPy was run on the same server, with QMCPy's UM-Bridge integration set to 36 parallel evaluations. This integration is a thin wrapper of QMCPy's existing functionality, and can issue parallel requests through Python's \texttt{multiprocessing} framework.

Online model evaluations average at around 20 minutes of run time, with most runs between 10 and 30 minutes. The main reason for strong fluctuations is that the number of subdomains that need to be recomputed varies between one and around eight, depending on defect location. We observe a total run time of 142m26s for 256 \gls{QMC} samples, indicating near-perfect speedup as expected due to \gls{QMC} samples' independence.

The preparatory offline run computing the low-dimensional basis on the entire domain was conducted on bwForCluster Helix of Baden-Württemberg, Germany. On 384 processor cores it took 113m30s. Compared to computing such a full MS-GFEM solution, the online method contributes a speedup of roughly 2000 in addition to the \gls{UQ} method's parallel speedup.

Containerization turned out crucial in this application: The MS-GFEM method is complex to set up and multiple parallel instances on the same system might lead to conflicts, which is alleviated by containerization. In addition, sharing model containers between collaborators greatly accelerated development.

\subsection{Multilevel Delayed Acceptance for tsunami source inversion}
\label{sec:application_mlda}

\subsubsection{Problem description}

We model the propagation of the 2011 Tohoku tsunami by solving the shallow water equations with wetting in drying. For the numerical solution of the PDE, we apply an ADER-DG method implemented in the ExaHyPE framework \cite{Reinarz2020, Rannabauer2018}.
We work with:
\begin{itemize}
    \item a model featuring smoothed bathymetry data incorporating wetting and drying through the use of a finite volume subcell limiter with polynomial order $2$ and a total of $1.7 \cdot 10^5$ degrees of freedom ($2187$ spatial); and
    \item a model using fully resolved bathymetry data and a limited ADER DG scheme of polynomial degree $2$ with a total of $1.7\cdot 10^7$ degrees of freedom ($6561$ spatial).
\end{itemize}
Further details on the discretization and the model can be found in \cite{Seelinger2021}, the model container is available in the UM-Bridge benchmark library.

\subsubsection{UQ workflow}

The aim is to obtain the parameters describing the initial displacements in bathymetry leading to the tsunami from the data of two DART buoys located near the Japanese coast\footnote{Bathymetry data was obtained from GEBCO \url{https://www.gebco.net/data_and_products/gridded_bathymetry_data/} and the DART buoy data for buoys 21418 and 21419 was obtained from NDBC \url{https://www.ndbc.noaa.gov/}.}. 

\begin{figure}[htbp]
    \centering
    \includegraphics[width=0.49\linewidth]{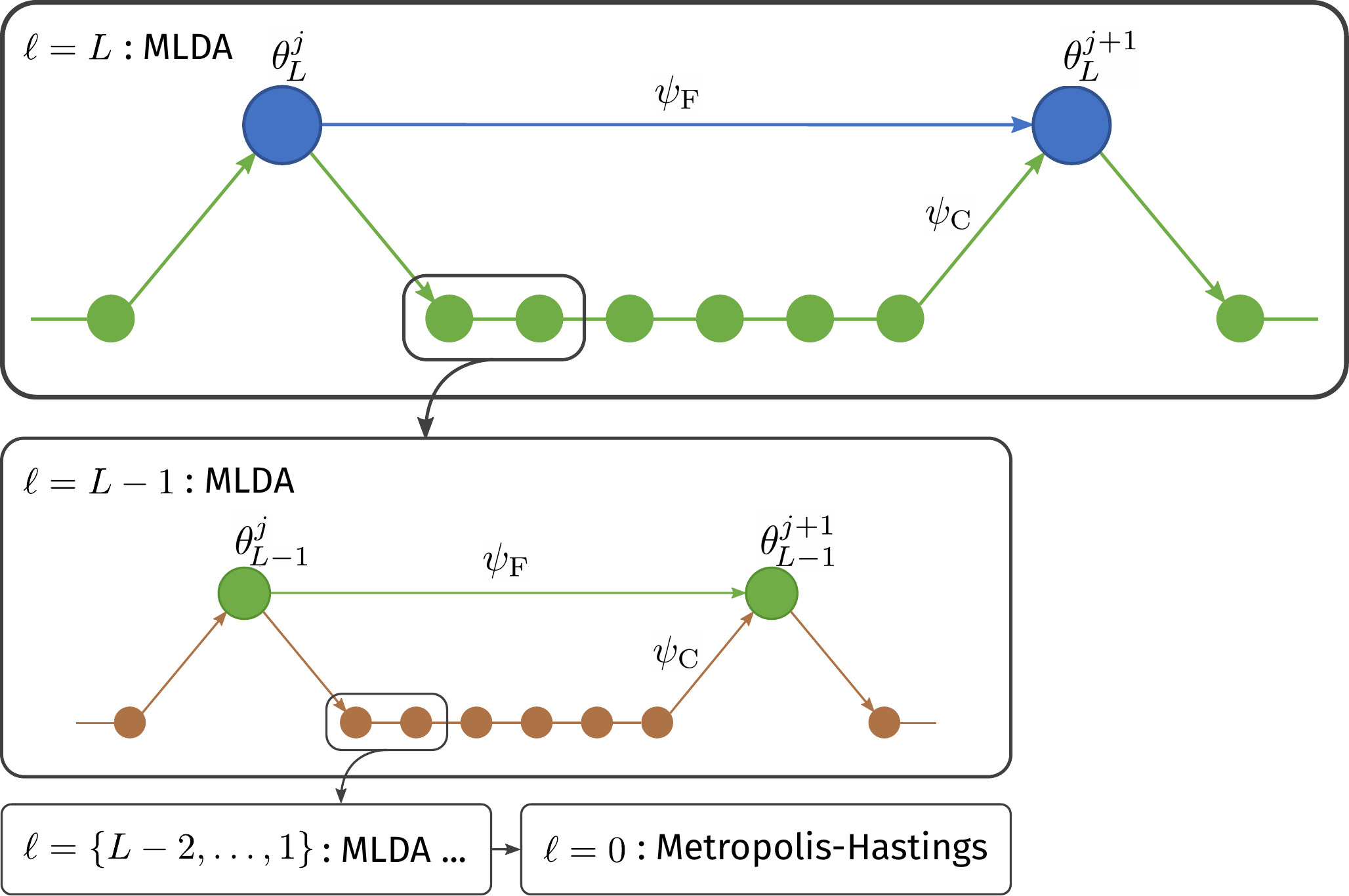}
    \includegraphics[width=0.49\linewidth]{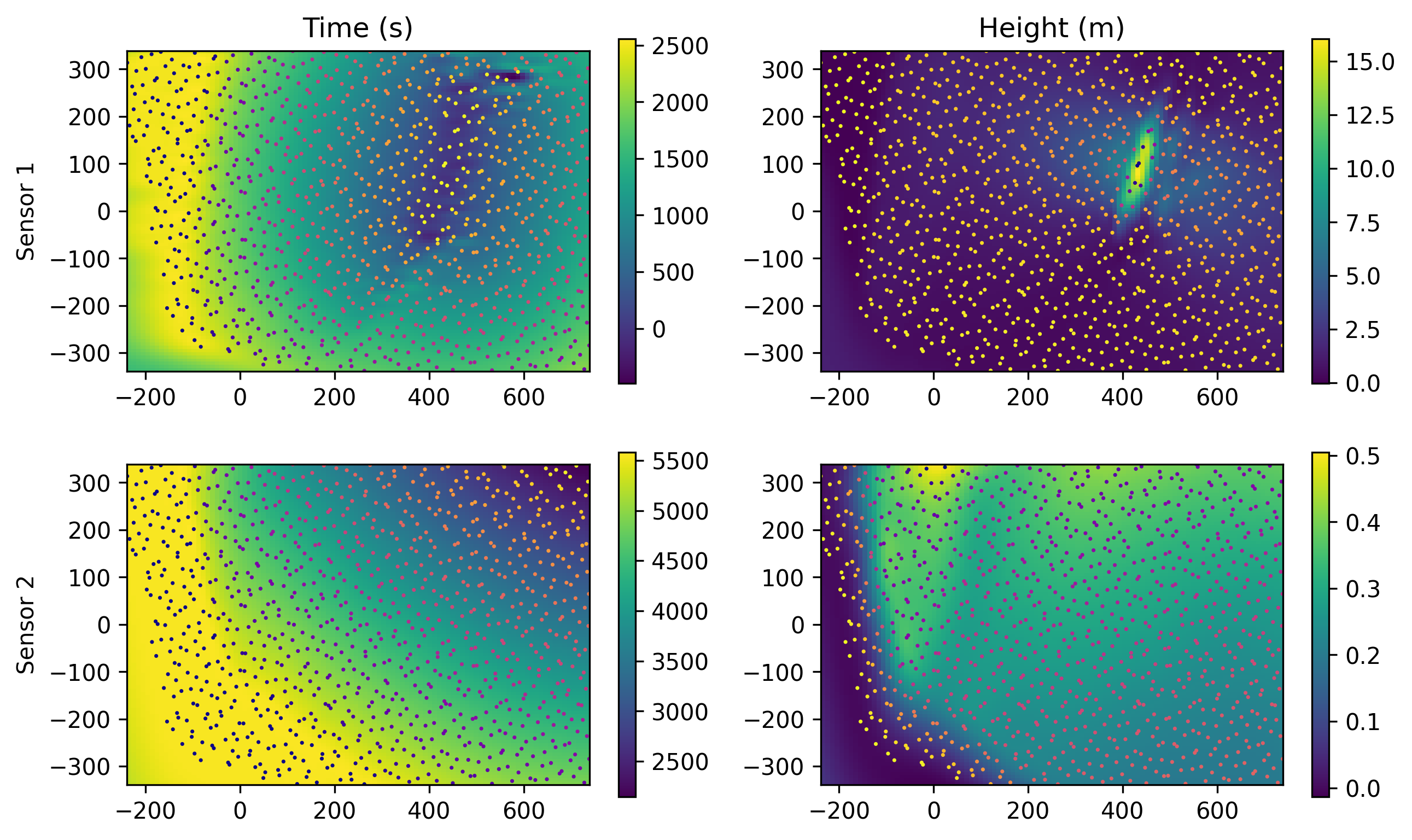}
    \caption{Left: GP training points (dots) and GP predictions (surfaces), for the arrival time and wave height for each of the sensors considered in the tsunami model.
    Right: The MLDA algorithm. Each level above the coarsest ($\ell=1 \dots L$) is sampled by recursively employing Delayed Acceptance (DA) MCMC. On the coarsest level ($\ell=0$), any Metropolis-Hasting algorithm can be used.}
    \label{fig:mlda}
\end{figure}

The posterior distribution of model inputs, i.e. the source location of the tsunami, was sampled with Multilevel Delayed Acceptance (MLDA) MCMC \cite{Lykkegaard_MLDA, lykkegaard_multilevel_2023} using the free and open-source Python package tinyDA \cite{tinyDA}. MLDA is a scalable MCMC algorithm, which can broadly be understood as a hybrid of Delayed Acceptance (DA) MCMC \cite{christen_markov_2005} and Multilevel MCMC \cite{dodwell_multilevel_2015}. The MLDA algorithm works by recursively applying DA to a model hierarchy of arbitrary depth, see \Cref{fig:mlda}

We used a three-level model hierarchy consisting of the fully resolved model on the finest level, the smoothed model on the intermediate level and a Gaussian Process emulator on the coarsest level. The \gls{GP} emulator was trained on $1024$ low-discrepancy samples of the model response from the smoothed model. The \gls{GP} emulator employed a constant mean function, a Matérn-$\frac{5}{2}$ covariance function with Automatic Relevance Determination (ARD) and a noise-free Gaussian likelihood. The \gls{GP} hyperparameters were optimized using Type-II Maximum Likelihood Estimation \cite{rasmussen_gaussian_2006}. Figure \ref{fig:mlda} shows the training samples and resulting \gls{GP} predictions for each sensor, as a function of the source location.

\begin{figure}[htbp]
    \centering
    \includegraphics[width=\linewidth]{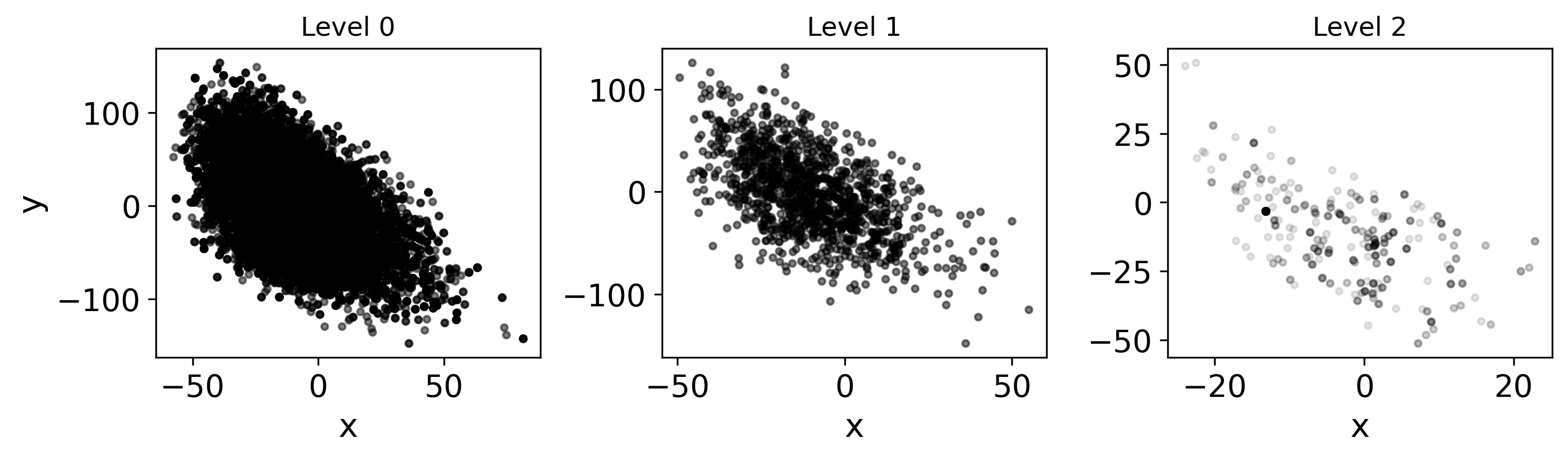}
    \caption{MLDA samples from each level, showing the posterior distribution of the source location of the tsunami.}
    \label{fig:tsunami_posterior}
\end{figure}

We employed a Gaussian Random Walk proposal, which had been pre-tuned to the posterior covariance induced by the \gls{GP} on the coarse level, to generate proposals. We set the subsampling rates for the coarse and intermediate levels to $25$ and $2$, respectively, resulting in near-independent samples passed to the finest level. We drew $7$ samples on the fine level for each independent chain, with $100$ independent, parallel chains. \Cref{fig:tsunami_posterior} shows the posterior samples from each level. While the coarse and intermediate levels exhibit very good mixing, and near-independent successive samples, the sampler on the fine level has over-sampled the initial point ($\mathbf{x}_0 = (-13, -3.5)$) and has not fully converged to the posterior distribution in this experiment. This bias is due to model discrepancy between the intermediate and fine levels and would be alleviated by running the MLDA samplers for longer. We consider this strong scaling issue of \gls{MCMC} methods out of scope for this work.

\subsubsection{Computational aspects}

\begin{figure}
    \centering
    \includegraphics[width=0.8\linewidth]{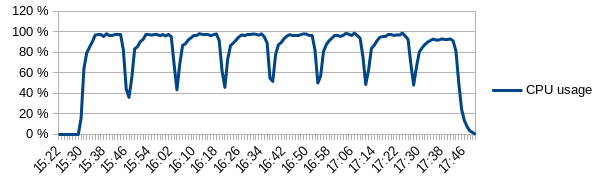}
    \caption{CPU usage on \gls{GKE} kubernetes cluster during MLDA run. Intermittent drops are caused by runs of the intermediate model, which require fewer resources.}
    \label{fig:mlda_cpu_usage_k8s}
\end{figure}

The model was run using the kubernetes setup of \Cref{sec:kubernetes} on \gls{GKE} with 100 \texttt{c2d-highcpu-56} nodes running 100 instances of the model container. In total, $2800$ physical CPU cores were used (counted as to $5600$ vCPUs including \gls{SMT} threads on \gls{GKE}). Each model instance was run on one \texttt{c2d-highcpu-56} node exploiting the Intel TBB parallelization capabilities of ExaHyPE \cite{Reinarz2020}.

Parallelization of the \gls{UQ} method was achieved through $100$ independent MLDA samplers. tinyDA allows running multiple samplers by internally making use of Ray \cite{liaw2018tune}, a Python library for distributed computing. The \gls{UQ} software was run on a workstation, connecting to the cluster above via tinyDA's UM-Bridge integration. Gaussian process evaluations were conducted directly on the workstation due to their low computational cost, while model evaluations were offloaded to the cluster.

Overall run time was 2h20min, with very effective use of the cluster as indicated by its CPU utilization in \Cref{fig:mlda_cpu_usage_k8s}. For parameter $(0,0)$, the smoothed model takes 1m1s to evaluate, while the fully resolved model takes 15m4s. We computed a total of 800 evaluations of the fully resolved model and 1400 of the smoothed model. Disregarding the additional inexpensive \gls{GP} evaluations, we obtain a parallel speedup of 96.38, giving a close to perfect speedup for 100 MLDA chains.

%% file: conclusion.tex
\section{Conclusion and future work}

We have demonstrated that UM-Bridge, a new universal interface between models and methods, facilitates UQ in cases ranging from small examples all the way to HPC-scale applications on cloud clusters. Neither models nor UQ software need to be adapted when scaling up, since most technical challenges are solved by the UM-Bridge interface and the associated kubernetes configuration. We have demonstrated the effectiveness of our approach on a number of realistic applications.

We hope that UM-Bridge will serve as a catalyst for many future UQ applications since it lowers the entry bar for UQ in general and for HPC-scale UQ in particular.

As a subsequent project, we aim to make the architecture of \cref{sec:kubernetes} available on HPC systems with SLURM \cite{SLURM} job scheduling and Apptainer \cite{singularity} support as well. To that end, we will develop a custom load balancer component corresponding to the role of HAProxy in the setting above. While it will not reach the same level of convenience, the same level of abstraction will be achieved regarding UQ and model software.